\documentclass[aps,prl,twocolumn,groupedaddress]{revtex4}
\usepackage{graphicx}
\usepackage{amsmath}
\usepackage[usenames]{color}

\begin{document}

\newcommand{\tco}{T$_{\mbox{\scriptsize CO}}$}
\newcommand{\tc}{T$_{\mbox{\scriptsize C}}$}

\newcommand{\LPSMO}{(La$_{\mbox{\scriptsize 1-z}}$Pr$_{\mbox{\scriptsize z}}$)$_{\mbox{\scriptsize 2-2x}}$Sr$_{\mbox{\scriptsize 1+2x}}$Mn$_{\mbox{\scriptsize 2}}$O$_{\mbox{\scriptsize 7}}$}
\newcommand{\LPSMOz}{(La$_{\mbox{\scriptsize 1-z}}$Pr$_{\mbox{\scriptsize z}}$)$_{\mbox{\scriptsize 1.2}}$Sr$_{\mbox{\scriptsize 1.8}}$Mn$_{\mbox{\scriptsize 2}}$O$_{\mbox{\scriptsize 7}}$}

\newcommand{\LSMO}{La$_{\mbox{\scriptsize 2-2x}}$Sr$_{\mbox{\scriptsize 1+2x}}$Mn$_{\mbox{\scriptsize 2}}$O$_{\mbox{\scriptsize 7}}$}
\newcommand{\LSMOb}{La$_{\mbox{\scriptsize 1.2}}$Sr$_{\mbox{\scriptsize 1.8}}$Mn$_{\mbox{\scriptsize 2}}$O$_{\mbox{\scriptsize 7}}$}
\newcommand{\oo}{ $(\frac{1}{4}\,\frac{1}{4}\,0)$}
\newcommand{\slp}{\mbox{(1/2,\,0,\,1/4$\pm \tau$)}}
\newcommand{\Akw}{$\mathcal{A}(\mathbf{k},\omega)$}

\newcommand{\jg}[1]{\textcolor{blue}{#1}}

\title{Electronic confinement and ordering instabilities in colossal magnetoresistive\\ bilayer manganites}

\author{J.~Trinckauf$^1$}
\author{T.~H\"anke$^1$}
\author{V.~Zabolotnyy$^1$}
\author{T.~Ritschel$^1$}
\author{M.O.~Apostu$^2$}
\author{R.~Suryanarayanan$^{3}$}
\author{A.~Revcolevschi$^3$}
\author{K.~Koepernik$^1$}
\author{T.K.~Kim$^1$}
\author{M.~v.~Zimmermann$^4$}
\author{S.V.~Borisenko$^1$}
\author{M.~Knupfer$^1$}
\author{B.~B\"uchner$^1$}
\author{J.~Geck$^1$}
\affiliation{$^1$Leibniz Institute for Solid State and Materials Research IFW Dresden, Helmholtzstrasse 20, 01069 Dresden, Germany}
\affiliation{$^2$Alexandru Ioan Cuza University of Iasi, 11th Carol I Blvd., 700506 Iasi, Romania}
\affiliation{$^3$LPCES - ICMMO - B\^at 410 Universit\'e Paris-Sud XI, 15, rue Georges Cl\'emenceau, 91405 Orsay Cedex, FRANCE}
\affiliation{$^4$HASYLAB at DESY, Notkestr. 85, 22603 Hamburg, Germany}

\date{Received: \today}

\begin{abstract}
	We present angle-resolved photoemission studies of \LPSMO\/ with $x=0.4$ and $z=0.1,0.2$ and $0.4$ along with density functional theory calculations and x-ray scattering data. Our results show that the bilayer splitting in the ferromagnetic metallic phase of these materials is small, if not completely absent. The charge carriers are therefore confined to a single MnO$_2$-layer, which in turn results in a strongly nested Fermi surface. In addition to this, the spectral function also displays clear signatures of an electronic ordering instability well below the Fermi level. The increase of the corresponding interaction strength with $z$ and its magnitude of $\sim$400\,meV, make the coupling to a bare phonon highly unlikely. Instead we conclude that fluctuating order, involving electronic and lattice degrees of freedom, cause the observed renormalisation of the spectral features.  
\end{abstract}


\maketitle

An important issue in the field of correlated electron physics concerns the relationship of collective charge carrier dynamics and electronic ordering tendencies\,\cite{Fradkin08012010}. In the limiting case where the interactions between the electrons are weak and the kinetic energy dominates, the resulting many-particle system can be described as a metallic Fermi-liquid of non-interacting quasiparticles. This changes dramatically, if the interactions between the electrons are strong and the dominant role of the kinetic energy is lost. In this limit, the charge carriers can crystallize, resulting in insulating behavior and static spatial modulations of the electronic system. In contrast to the Fermi-liquid, these electronic crystals typically break the symmetry of the underlying nuclear lattice. 

A most intriguing case might be realized just in between these two limits, where neither the interactions between the electrons nor the kinetic energy is dominating. In these cases, electronic liquid crystals may emerge, where the electrons crystallize to some extend, while still forming a conductive phase\,\cite{FradkinReview2010}. Electronic liquid crystals, and more specifically nematic electronic order, are currently discussed intensively in relation to the unconventional superconductivity and the pseudo-gap phase of the high-temperature superconducting cuprates\,\cite{Lawler2010,Kivelson1998,HinkovNature2008}. There are also clear indications for such an ordering in some of the new iron pnictide superconductors\,\cite{Chuang08012010}. 

The observed proximity of metallic and ordered phases in doped manganites shows that also in these materials ordering tendencies and kinetic energy both are important and  competing\,\cite{Tokura2000,TokuraRepProgPhys06}. It is also widely believed that this competition lies at the heart of the famous colossal magnetoresistance (CMR) effect, but the detailed physical mechanisms driving the CMR still remain to be fully understood. 
In the bilayer CMR-manganites  \LSMOb \/ ($x=0.4$), the competition between electronic ordering tendencies and charge delocalization is particularly prominent: First, the growth of ordered regions with cooling in the paramagnetic insulating phase (PMI) is observed in scattering experiments\,\cite{PhysRevLett.83.4393}. Then, when the transition into the ferromagnetic metallic (FMM) phase is reached at \tc, these ordered regions melt and the CMR occurs\,\cite{PhysRevLett.83.4393,PhysRevB.67.104433}. 
Correspondingly, unconventional electronic properties of the FMM phase were indeed observed in previous angle-resolved photoemission spectroscopy (ARPES) studies, including a strongly doping and temperature dependent pseudo-gap\,\cite{Chuang2001,Mannella2005, PhysRevB.76.233102}.

In this letter, we present a study of \LPSMOz\/ (LPSMO) with a fixed hole concentration of $x=0.4$. Previous macroscopic studies of these samples showed that the FMM phase is strongly destabilized with increasing Pr-concentration $z$\,\cite{PhysRevB.67.104433}. The reduction of \tc\/ with increasing $z$ indicates that the chemical pressure due to the smaller Pr affects the balance between kinetic energy and electronic ordering tendencies. 
Here, we apply ARPES to investigate how altering this balance changes the electronic structure and the charge dynamics of LPSMO. ARPES provides direct access to the single particle spectral function \Akw\/\,\cite{RevModPhys.75.473} and, in combination with density functional theory calculations and x-ray diffraction data, enables us to reveal the effects of fluctuating electronic order on the charge dynamics in the FMM phase of LPSMO.

The studied LPSMO single crystals were grown by the traveling floating zone method. It was found that a small amount of Pr enables to grow single crystals of very high quality as confirmed by x-ray diffraction, magnetization and thermal transport measurements. The ARPES measurements were performed using the $1^3$-station at the  beamline UE112 of the synchrotron facility BESSYII, which provides a beamspot of about $100~\mu m \times 100~\mu m$. All presented ARPES spectra were measured with 55\,eV excitation energy at around 23~K, if not stated otherwise. The diffuse x-ray diffraction experiment was performed at the high-energy beamline BW5 at HASYLAB in Hamburg. At the used photon energy of 100\,keV, the penetration depth is of the order of 1\,mm, guaranteeing the detection of true bulk properties. The density functional theory calculations in the local spin density approximation (LSDA) and the subsequent tight-binding analysis were performed using the code FPLO 8.5\,\cite{PhysRevB.59.1743}. All calculations for the hole doping of $x=0.4$ were done in the virtual crystal approximation.

\begin{figure}
  \includegraphics[width=9.5cm]{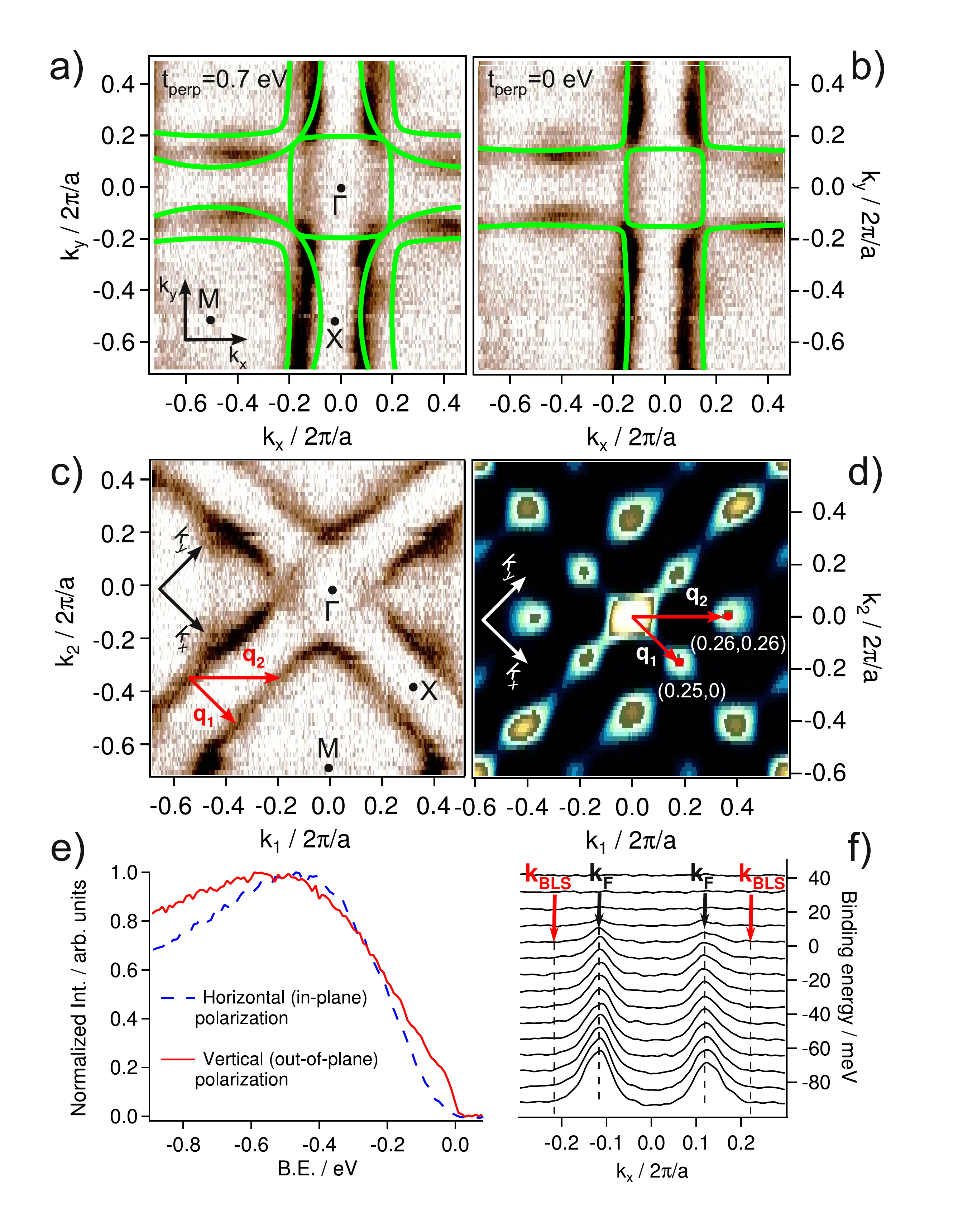}
  \caption{Fermi surface (FS) of (La$_{0.9}$Pr$_{0.1}$)$_{1.2}$Sr$_{1.8}$Mn$_2$O$_7$ measured at 23~K with out-of-plane polarization. The green lines in a) and b) show the calculated FS with the full LSDA interlayer hopping $t_{\mathrm{perp}}=0.7$\,eV and $t_{\mathrm{perp}}=0$\,eV, respectively. c) FS map measured with the sample rotated by 45$^{\circ}$ and out-of-plane polarization. d) The autocorrelation function of the FS map in c).  ${\mathbf q}_1$ and ${\mathbf q}_2$ indicate nesting vectors. The spectra have been intergrated $\pm40$~meV around the Fermi level. e) EDC integrated several meV around $\mathbf{k}_F$ for in-plane and out-of-plane polarization normalized to its maximum. The relative spectral weight of the features near $E_F$ is much bigger with out-of-plane polarization, revealing a small Fermi edge showing that the samples are indeed metallic. These states make up the FS maps in a) - c). f) MDC of a spectra parallel to $\mathbf{k}_x$ at $\mathbf{k}_y$ around $\mathbf{k}=(k_x,-0.7)$. No BLS can be seen even though the TB model for the FMM phase predicts a band splitting of around 0.1~$2\pi/a$. \label{FSvsWannier}}
\end{figure}

Figures\,\ref{FSvsWannier}\,a) - c) show the measured Fermi surface (FS) of LPSMO with $z=0.1$. Very similar results were obtained for $z=0.4$. These data were taken deep inside the FMM phase (\tc$\approx$ 100~K) and the polarization was chosen as to enhance the spectral weight at the Fermi level, as demonstrated in Fig.\,\ref{FSvsWannier}\,e). The EDCs in Fig.\,\ref{FSvsWannier}\,e) also show that the spectral weight at the Fermi energy $E_F$ is small, which is in line with the small Drude weight observed for these compounds\,\cite{Kimura2010}. The measured remnant FS formed by these states consists of parallel straight sections over large regions of $\mathbf{k}$-space, which agrees with previous ARPES studies. This well-nested FS already indicates the presence of electronic ordering instabilities. 

According to LSDA, there is a sizable hybridization between the two planes within a bilayer, which is reflected by the large inter-plane hoping $t_{perp}=0.7$\,eV obtained from a TB-Wannier-analysis. This large $t_{perp}$ leads to a considerable splitting of bonding and anti-bonding bands; the so-called bilayer splitting (BLS). Along the M--X--M direction, for instance, the calculated BLS at $E_F$ is about 0.1$\times2\pi/a$.
As can be seen in Fig.\,\ref{FSvsWannier}\,a), 
this does not agree with experiment, where such a big BLS is not observed. In fact, the data displayed in Fig.\,\ref{FSvsWannier}\,f), clearly excludes the presence of a large BLS in LPSMO: even though the momentum distribution curves (MDCs) close to $E_F$ are broad, their symmetric shape clearly shows that the BLS in the real system is much smaller, if not completely absent. This result was reproduced with different polarizations and in different regions in $\mathbf{k}$-space, excluding possible matrix element effects.
The reduction of the BLS observed here is in perfect agreement with the results of Jozwiak {\it et al.}, who also discovered a dramatic reduction of the BLS for the Pr-free compound between x=0.36 and 0.4\,\cite{PhysRevB.80.235111}. In addition, very similar FS were observed in single-layer manganites, where a BLS does not exist\, \cite{Evtushinsky2010}.

Within our TB-model it is possible to set $t_{perp}=0$\,eV, i.e., to switch off the bilayer hybridization and to eliminate the BLS. As can be seen in Fig.\,\ref{FSvsWannier}\,b), this gives a much better agreement with experiment, further supporting the absence of the BLS. The FS of the real material is therefore very close to the independent particle reference given by this TB-model. Importantly, the TB-results given in Fig.\,\ref{FSvsWannier}\,b) imply that setting $t_{perp}=0$\,eV yields a strongly nested FS with large parallel sections, which is indicative for electronic ordering instabilities\,\cite{Gruener_Book}. In other words, the confinement of the charge carriers to a single MnO$_2$-plane strongly enhances electronic ordering tendencies.

The fact that the measured FS is well described by an unreconstructed TB band structure, motivates to determine the so-called nesting vectors from the experimental data. These vectors, which indicate possible wave vectors for the electronic ordering, can be determined by calculating the autocorrelation function of the measured FS. This is shown for a specific example in Figs.\,\ref{FSvsWannier}\,c) and d), where the experimental geometry was chosen as to yield similar photoelectron intensities in the different regions of $k$-space. Equivalent autocorrelation functions were obtained for all other measurements as well. The peaks in the autocorrelation function correspond to ${\mathbf q}_1=(0.25,0)$ and ${\mathbf q}_2=(0.26,0.26)$ with an estimated error of $\Delta q=\pm(0.01,0.01)$ (in units of $2\pi/a$). Note that ${\mathbf q}_{2}$ corresponds closely to the orbital ordering in the so-called CE-order of half doped manganites\,\cite{Tokura2000}. Further, two inequivalent nesting vectors indicate that the observed FS supports two different electronic ordering instabilities. 


\begin{figure}
  \includegraphics[width=8.8cm]{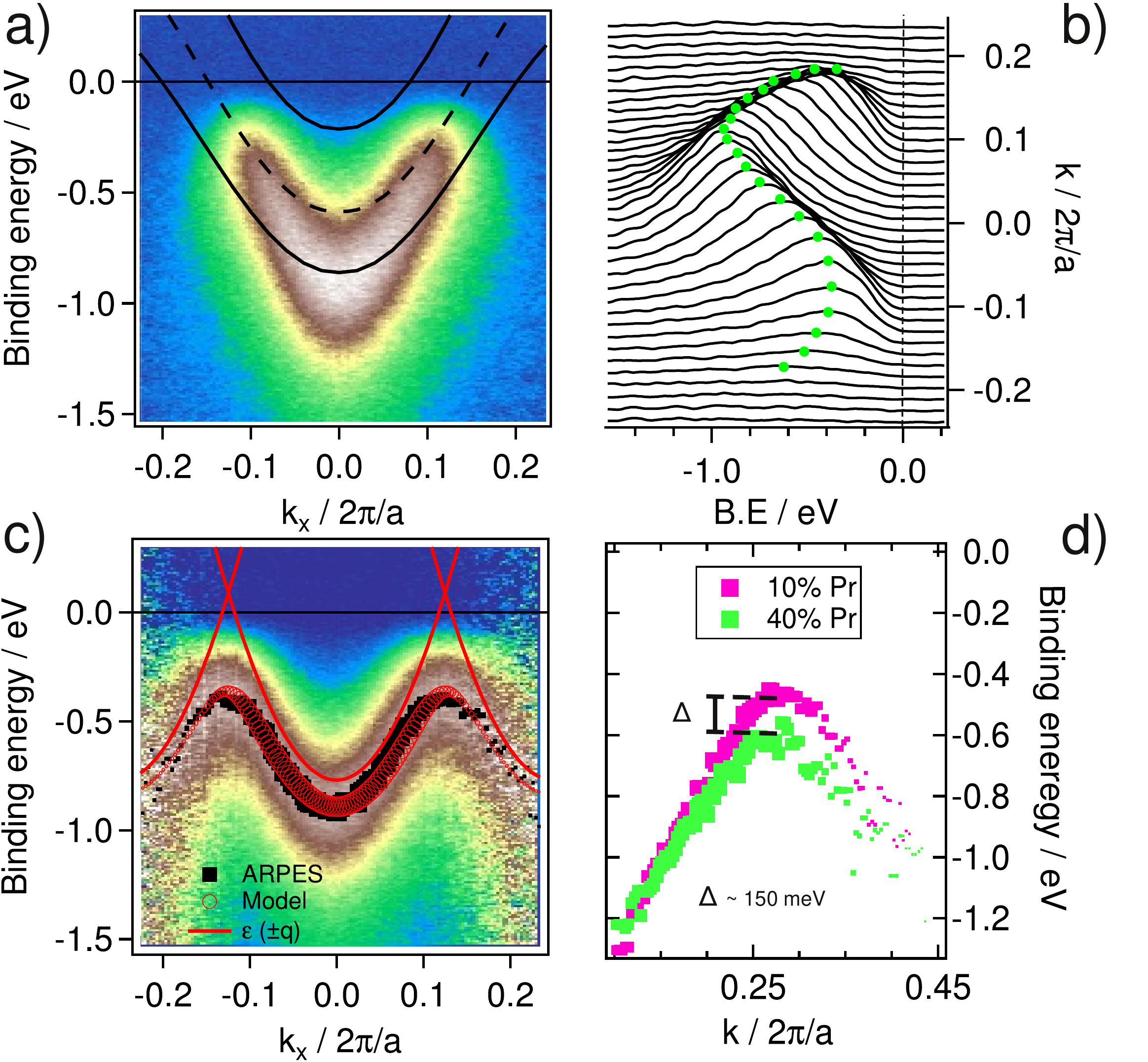}\\
  \caption{a) Measured spectrum of (La$_{0.9}$Pr$_{0.1}$)$_{1.2}$Sr$_{1.8}$Mn$_2$O$_7$ along $M-X-M$ after subtraction of a nondispersive background feature compared to the TB-bands for a BLS of 0.7~eV (black solid line) and no BLS (black dashed line).  b) Constant momentum plot of the spectrum shown in a). The dots are a guide to the eye emphasising the back-bending of the momentum dispersion. c) Dispersion of the EDC maxima of spectrum a) (black squares) compared to a fitted model function (red open circles). Marker size represents photoemission/spectral intensity. The false-color plot shows the same spectra as in a) and b) but each EDC is normalized to its maximum to enhance states with weak intensity. The back-bending is clearly observable. d) Comparison of the gap size for $z=0.1$ and $z=0.4$. Dispersion is taken from the EDC maxima of spectra measured along $\Gamma-M$.  All spectra were measured with in-plane polarization at 23~K. \label{wf}}

\end{figure}

However, FS nesting alone does not necessarily imply that the electronic order at  ${\mathbf q}_{1,2}$ really occurs \cite{Johannes2008}. We therefore also investigated \Akw\/ well below Fermi level $E_F$. Figure\,\ref{wf}\,a) and b) show a measured spectrum along M--X--M. As can be seen Fig.\,\ref{wf}\,b), near the Fermi wave vector $\mathbf{k}_F$, the dispersion shows a  back-bending with quickly decaying spectral intensity. Such a back-bending is a clear fingerprint of electronic ordering. To demonstrate this, we follow the procedure outlined in  Refs.\,\onlinecite{Hashimoto2010} and \onlinecite{PhysRevB.77.235104} and calculate the spectral weight redistribution starting from the Hamiltonian
\begin{displaymath}
\mathcal{H}=\sum_\mathbf{k} \varepsilon(\mathbf{k}) c^{\dagger}_{\mathbf{k}}c_{\mathbf{k}} + \sum_{\mathbf{k},\mathbf{q}} g_\mathbf{q} c^{\dagger}_{\mathbf{k}+\mathbf{q}}c_{\mathbf{k}}(a_{\mathbf{q}}+a^{\dagger}_{\mathbf{-q}})\, ,
\end{displaymath}
which describes the coupling of quasiparticles to a collective bosonic mode. In the above equation, $\varepsilon_\mathbf{k}$ is the unperturbed single-particle band dispersion, $g_\mathbf{q}$ is the boson-electron coupling and $c^{\dagger}_{\mathbf{k}}$ ($c_{\mathbf{k}}$) and $a^{\dagger}_{\mathbf{q}}$ ($a_{\mathbf{q}}$) create (annihilate) a fermion or boson, respectively. 
As justified  by the observed quasi one-dimensional dispersion around the X-point, we set $\varepsilon(\mathbf{k}_x)=a+b\cos{\mathbf{k}_x}$. Further, we assume a single modulation with wave vector $\mathbf{q}$, treat the bosonic mode on a mean field level and calculate the eigenfunctions and energies of $\mathcal{H}$. 
Using $a$, $b$, $\mathbf{q}_x$ and the interaction strength $V=2g_{\mathbf{q}} \langle a_{\mathbf{q}}\rangle$ as free parameters, we fit the model to the measured dispersion taken from the maxima of the energy distribution curves (EDCs) at each $\mathbf{k}_x$. The results given in  Fig\,\ref{wf}\,c) show an excellent agreement between the measured and the modelled dispersion for $\mathbf{q}_x=0.25$ and $V=432$\,meV.
Note that the fitted odering wave vector $\mathbf{q}_x=0.25$ is fully consistent with the measured nesting vectors $\mathbf{q}_{1,2}$. 
It is also important to realize that, although the spectral intensity is not fitted to the data, the model reproduces  $\mathbf{q}_x$-dependence of the measured peak intensity extremely well. 
The very good agreement of experiment and model shows that \Akw\/ in the FMM phase displays clear signatures of electronic ordering at a wave vector that agrees with the FS nesting vectors. 

\begin{figure}
  \includegraphics[width=0.98\columnwidth]{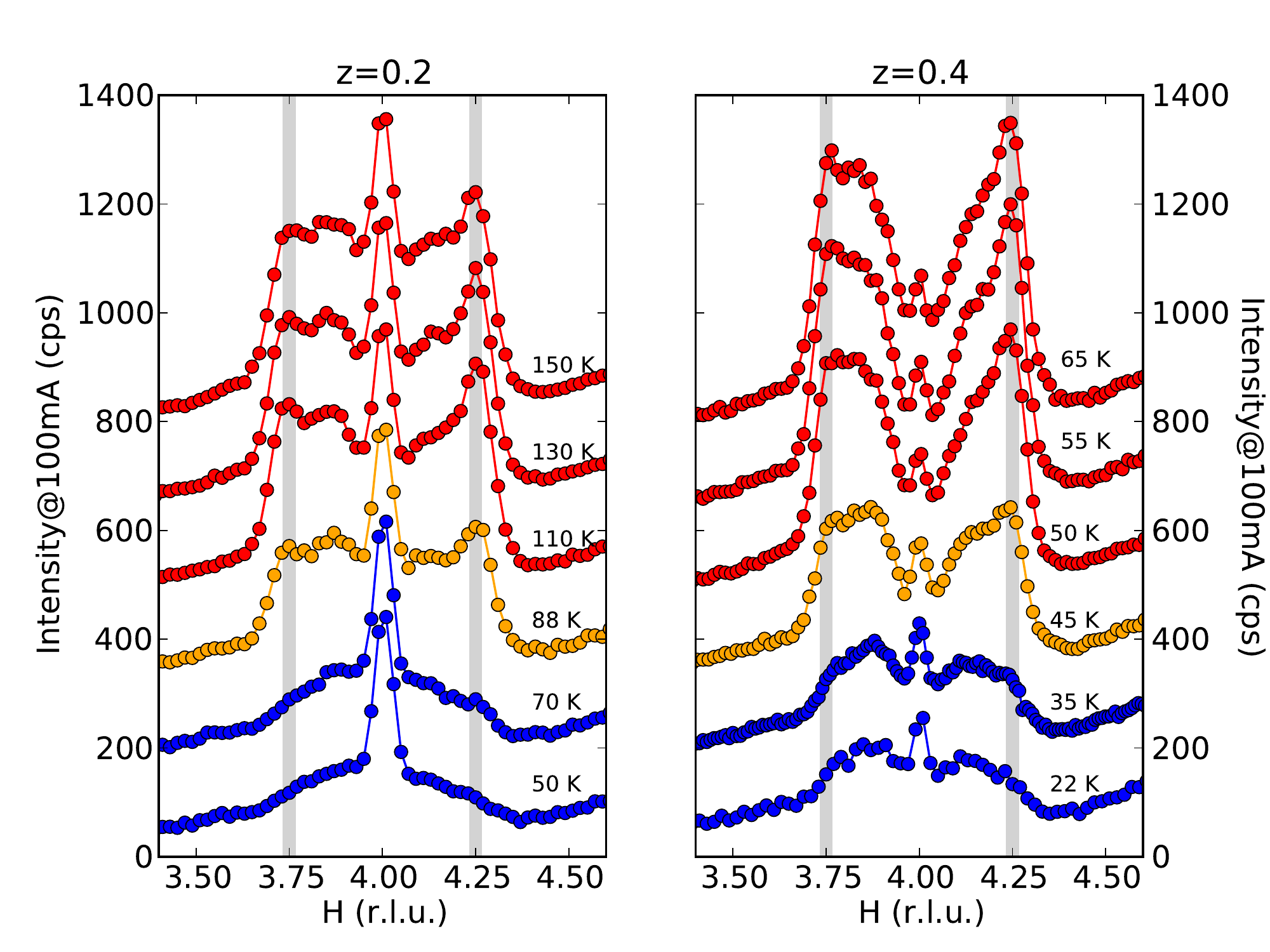}\\
  \caption{X-ray diffuse scattering data of \LPSMOz\/ with $z=0.2$ and $z=0.4$ for different temperatures around \tc. The weak symmetry forbidden (4,0,1) Bragg peak is due to a small amount of stacking faults. Therefore, the (401) intensities of the two samples cannot directly be compared. Above the Curie temperature of around 90\,K ($z=0.2$) and 45\,K ($z=0.4$) broad superstructure reflections at $(4\pm0.25,0,1)$ are observed. The data sets at 88\,K ($z=0.2$) and 45\,K ($z=0.4$) were taken just above \tc. The ordered regions melt upon entering the FMM phase.} \label{xrd}
\end{figure}

In order to study possible static ordering in our samples, we performed elastic high-energy x-ray diffraction (HE-XRD) studies. Figure~\ref{xrd} shows the data obtained for samples with $z=0.2$ and $z=0.4$ at various temperatures around \tc . Referring to the space group I4/mmm, weak intensity was detected at the symmetry forbidden (401) reflection due to a small amount of stacking faults within the probed sample volume. 

More importantly, strong diffuse scattering intensity was observed around the (401). At room temperature, only broad diffuse scattering is observed (not shown) which has been attributed to uncorrelated polaronic lattice distortions \,\cite{PhysRevLett.83.4393}.  Upon cooling down towards \tc , superlattice peaks centered at $(4\pm0.25,0,1)$ emerge out of this broad diffuse scattering, signaling the growth of ordered regions within the disordered majority phase. With entering the FMM phase at \tc , these superlattice peaks vanish abruptly and only diffuse scattering remains in the FMM phase. We also verified the absence of static CE-order below \tc \/ in our samples.  Similar diffuse scattering results were obtained for La$_{1.2}$Sr$_{1.8}$Mn$_2$O$_7$, where the temperature dependence of the broad superlattice reflections was discussed in terms of ordered areas with a correlation length of about 27\,\AA, which melt upon entering the FMM phase \,\cite{PhysRevLett.83.4393}. Interestingly, the modulation vector of (0.25,0,0) observed by HE-XRD agrees perfectly with the nesting vector $\mathbf{q}_1$ defined by the FS topology. This agreement between the FS topology, the back-bending and the measured ordering strongly argues in favor of ARPES representing bulk properties. 

We note that the modulation vector (0.25,0,0) found for LPSMO differs from the (0.3,0,0) observed for La$_{1.2}$Sr$_{1.8}$Mn$_2$O$_7$\,\cite{PhysRevLett.83.4393}. The reason for this difference is currently unclear. However, we can already exclude changes in the doping level caused by Pr, because the modulation vector determined by HE-XRD for LPSMO is the same for $z=0.2$ and $0.4$. In addition, the FSs measured for $z=0.1$ and $z=0.4$ show no sign of a difference in hole doping. This interesting point is subject of current investigations, but plays no role for the arguments presented in this paper.

To summarize, the main result of the present ARPES study is the observation of clear fingerprints of electronic order deep inside the FMM phase of LPSMO. Apart from the strongly nested FS, which  has also been reported previously for other bilayer compounds\,\cite{Sun2009a}, we identified the presence of a back-bending anomaly in the measured \Akw. The excellent agreement between experiment and the presented model calculation establishes the coupling of the charge carriers to a collective excitation, which causes an inherent breaking of translational symmetry. The large coupling strength $V=432$\,eV found for $z=0.1$ makes a coupling to a bare phonon, as in the case of a conventional CDW transition, highly unlikely. This is further supported by the observed destabilisation of the FMM phase with Pr doping \cite{PhysRevB.78.060406}, which is consistent with the measured increase in gap-size (see Fig.\,\ref{wf}\,d)) and points towards an electronic origin of the underlying ordering phenomenon. At the same time, HE-XRD shows no evidence for static order in the FMM phase. We therefore conclude that fluctuating order exists in the FMM phase, which has a strong impact on the charge carrier dynamics. Such an order is expected to involve all electronic and lattice degrees of freedom, which is a peculiarity of the mixed valence manganites.

A likely and consistent scenario are strong fluctuations between a two-dimensional metallic phase and the ordered CE-phase, which both favour the occupation of in-plane orbitals. In fact, evidence for CE-fluctuations in the FMM phase of \LSMOb\ was recently deduced from inelastic neutron scattering data\,\cite{Weber2009}. In addition, the TB-analyis presented above clearly indicates that a two-dimensional metallic phase of LPSMO is indeed unstable towards electronic order. The notion of fluctuating CE-order in the FMM phase close to $x=0.5$ is therefore well supported by experimental and theoretical results. In particular, this scenario naturally explains the observed coexistence of a metallic FS and the back-bending anomaly reported here. 

Finally we note that the FMM phase exhibits typical features of an electronic liquid crystal: (i) proximity to static order (diffuse scattering, FS nesting), (ii) evidence for molten electronic order (back-bending, no static order), and (iii) reduced electronic dimensionality (FS, BLS). This strongly motivates further studies of CMR materials in this context.

{\it Acknowledgments} We thank S. Leger, R. Sch\"onfelder and R. H\"ubel for the technical support. JG, TR and JT gratefully acknowledge the financial support by the DFG through the Emmy Noether Program (Grant GE1647/2-1).

\bibliography{manuscript}
\bibliographystyle{apsrev}

\end{document}